\documentclass[prl,aps,showpacs,twocolumn,preprintnumbers,amsmath,amssymb,superscriptaddress,longbibliography]{revtex4-1}
\usepackage[english]{babel}
\usepackage{amsmath,amssymb,bbm,mathrsfs,mathtools,multirow,diagbox,xcolor,
  graphicx,tabularx,comment,amsfonts,dsfont,lettrine,units,comment}
\usepackage{graphicx}
\usepackage[colorlinks=True,linkcolor=red,citecolor=blue,urlcolor=blue]{hyperref}
\usepackage{bookmark}
\usepackage{xcolor}
\usepackage{chngcntr}
\usepackage{braket}
\usepackage{nicefrac}
\usepackage{bm}
\usepackage{bbm}
\usepackage{braket}
\usepackage{lineno}
\usepackage{array}
\newcolumntype{C}{>{$}c<{$}}
\usepackage{newtxtext} 
\usepackage{ulem}   

\renewcommand{\dag}{^{\dagger}}

\def\br{\mathbf{r}}
\def\bk{\mathbf{k}}

\def\ket#1{|#1\rangle }
\def\bra#1{\langle #1 |}

\usepackage{xstring} 

\begin{document}

\title{Obstructed insulators and flat bands in topological phase-change materials
}

\author{Quentin Marsal}
 \affiliation{Univ. Grenoble Alpes, CNRS, Grenoble INP, Institut N\'eel, 38000 Grenoble, France}

\author{Daniel Varjas}
\affiliation{Department of Physics, Stockholm University, AlbaNova University Center, 106 91 Stockholm, Sweden}

\author{Adolfo G. Grushin$^*$}
\affiliation{Univ. Grenoble Alpes, CNRS, Grenoble INP, Institut N\'eel, 38000 Grenoble, France}

\begin{abstract}
    Phase-change materials are ubiquitous in technology because of their ability to transition between amorphous and crystalline phases fast and reversibly, upon shining light or passing a current.  
    Here we argue that to fully understand their electronic properties it is necessary to define a novel electronic phase: the amorphous obstructed insulator.
    It differs from an obstructed insulator crystal in that it presents localized edge or surface states irrespective of the sample termination. 
    Consequently, we show that obstructed amorphous insulators in three-dimensions host a surface two-dimensional flat band, detectable using ARPES.
    Our work establishes basic models for materials where topological and obstructed properties can be switched on and off externally, including two-dimensional surface flatbands.
\end{abstract}

\date{\today}

\maketitle

Externally controlling the topological state of a material is an appealing challenge to integrate topological properties into technology~\cite{Gilbert2021}.
For example, temperature~\cite{Eremeev2016}, pressure~\cite{Xi2013,Ideue2014,Liang2017,Toshiya2019}, composition~\cite{Hsieh2008,Hsieh2009,Xu11,Brahlek2012,Dziawa2012} or magnetic fields~\cite{Veyrat20} can drive a topological transition between different crystalline states, but these are not always fast, practical or reversible processes. 

In this work we explore a different way to control the topological phase, based on the controllable atomic structure of phase change materials~\cite{Eremeev2016,Fantini_2020}. Phase-change materials transition from a crystal to an amorphous solid state reversibly, by means of an electrical current or light pulse. For example, a short and intense current melts the crystal, to later freeze it into an amorphous state, while a longer and weak current relaxes the latter back to a crystal. The two states, crystalline and amorphous, strongly differ in their electric resistance. This property positions phase-change materials at the center of commonplace optical memories, such as DVDs, and next-generation non-volatile random access memories~\cite{Fantini_2020}.

While the topological character of the crystalline phase of phase-change materials can be consulted in online databases (see Ref.~\cite{Wieder22} for a review) the investigation of topological properties of their amorphous phases is an ongoing experimental effort~\cite{Reindl2019,Korzhovska2020}. 
The task of theoretically understanding the fate of topological properties of the amorphous state of phase-change materials has recently become possible~\cite{Grushin2020}. Many phases familiar from crystals have been proposed in amorphous matter as well, such as Chern insulators~\cite{Agarwala:2017jv,Mitchell2018,Poyhonen2017,Bourne:2018jr,Marsal2020}, quantum-spin-Hall effects or three-dimensional (3D) topological insulators~\cite{Agarwala:2017jv}. Moreover, topological properties in realistic amorphous models can be signaled using symmetry indicators~\cite{Marsal2020}, that generalize those used in crystals~\cite{Kruthoff17,Po:2017ci,Song2018,Varjas2019}, as an additional tool to  real-space markers~\cite{LCM,Grushin2020}.

A notable absence is the notion of obstructed atomic insulators in the amorphous form. Obstructed phases~\cite{Song2017,Rhim2017,Bradlyn2017,Benalcazar2019,Schindler2019,Cano2022} can be described by localized wavefunctions, and thus are topologically trivial in a strict sense.
However, they can be classified by symmetry, through the quantization of a winding number, that restricts the average position of charge centers to be at high-symmetry points of the unit-cell, away from the atomic sites~\cite{Xu2021}.
Symmetry guarantees a gapless state between two obstructed insulators at specific electronic fillings. The half-filled Su-Schrieffer-Heeger (SSH) model~\cite{Su79} for a Peierls distorted (or dimerized) poly-acetylene chain is a well studied example, with higher-dimensional generalizations~\cite{Song2017,Rhim2017,Bradlyn2017,Benalcazar2019,Schindler2019,Cano2022}.
However, obstructed insulators have eluded their generalization to the amorphous state, because the unit cell symmetries seem necessary to define them.

The absence of a theory for amorphous obstructed insulators challenges our understanding of known phase-change materials. Two technologically relevant examples are GeTe and Sb$_2$Te$_3$. When crystalline, the former is a trivial insulator, and the latter is a 3D topological insulator protected by time-reversal symmetry~\cite{Eremeev2016,Vergniory:2019ub, Bradlyn2017, Vergniory2021, TopoQuantumChemistry, Aroyo2011, Aroyo:xo5013, Aroyo2006}. Structural analysis shows that several phase-change materials, including GeTe and Sb$_2$Te$_3$, exhibit an average dimerization in their amorphous form: bonds locally aligned with each other tend to alternate between weak and strong~\cite{Raty2015,muralidharan2022investigation}. In the amorphous state, this average dimerization can even increase over time, causing a resistance drift~\cite{Raty2015}. Hence, defining the possible obstructions relevant for dimerized amorphous phases seems necessary to understand phase-change materials. 

Here we argue that phase-change materials are a natural platform to realize controllably amorphous obstructed insulators, a phase that we define in this work. Compared to crystalline obstructions, amorphous realizations have the advantage that localized boundary states appear in all terminations. As a consequence, a filling-enforced two-dimensional (2D) flat band can appear and disappear controllably in 3D phase-change materials, like GeTe, by changing from its crystal to its amorphous form. Our goal is to propose models and experiments that display this controllable phenomenology, unique to amorphous matter.

\begin{figure}
    \centering
    \includegraphics[width = \columnwidth]{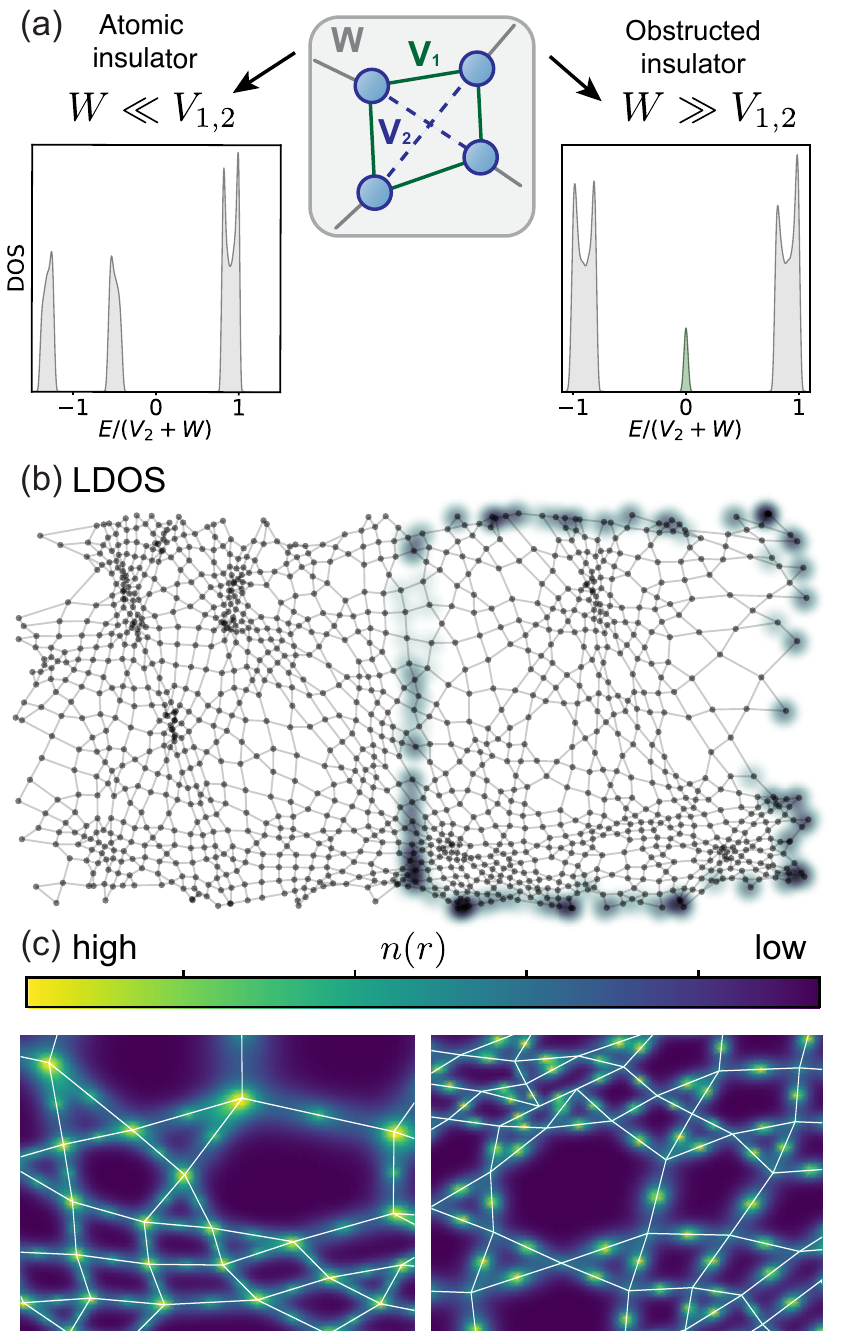}
    \caption{Amorphous obstructed and atomic insulators. (a) The center sketch shows the parameters of the Hamiltonian in Eq.~\eqref{eq:WT}. When $W$ is the smallest energy scale (left), the DOS shows a gap at half-filling. When $W$ is the largest energy scale, the DOS presents in-gap states (right). (b) The boundary between the obstructed insulator and the atomic insulator (or vacuum) shows localized in-gap states in the local DOS. (c) The charge centers of the filled states of the atomic insulator $n(r)$ show a strong maximum at the atomic sites (left), unlike the obstructed insulator, which shows a maximum between the atomic sites (right). The parameters are $(W,V_1,V_2)=(0.1,0.2,0.9)$ and $(W,V_1,V_2)=(0.9,0.01,0.1)$ for the atomic and obstructed insulators, respectively. 
    }
    \label{fig:spectrum}
\end{figure}

\textit{2D Amorphous obstructed insulators.}
In crystals, different obstructed atomic limits usually occur when the symmetries of the unit cell constrain the position of the charge centers to be at different symmetry centers. A 1D example is the SSH chain with inversion symmetry~\cite{Su79}, in which the charge centers of the occupied states are localized between two sites. Inversion symmetry obstructs localizing them at the site, which defines the atomic insulator limit.

Our first goal is to show that amorphous obstructed and atomic insulators can be also defined in amorphous lattices. 
In amorphous systems, lattice disorder breaks long-range translational invariance but preserves the local arrangement of the atoms~\cite{Zallen}. In other words, there is no single crystal from which the amorphous lattice is obtained, and hence no well defined disorder-free limit.
A historically fruitful approach to establish a tractable limit is to neglect fluctuations in the hopping or onsite terms of the Hamiltonian and assume that the disorder is only structural, i.e., induced by the random connectivity of the lattice~\cite{Weaire1970,Zallen}.
In particular, Weaire-Thorpe models~\cite{Weaire1970, Marsal2020} describe amorphous systems composed by a disordered arrangement of identical unit blocks. By construction they have fixed coordination, a realistic local property of amorphous materials~\cite{Zallen}, also useful to define amorphous obstructed atomic limits. To this end, consider a 2D Weaire-Thorpe Hamiltonian~\cite{Marsal2020}
\begin{equation}
    \mathcal{H} = \sum_i \mathbf{c}_i^{\dag} H_V\mathbf{c}_i + \sum_{\left<i,j\right>}\mathbf{c}_i^{\dag} H_W(i,j)\mathbf{c}_j,
    \label{eq:WT}
\end{equation}
where every site, shown at the center of Fig. \ref{fig:spectrum}(a), is fourfold coordinated. At a given site, $H_V$ couples its four orbitals, encoded in the creation and annihilation operators $\mathbf{c}_i^{\dag}$ and $\mathbf{c}_i$, respectively. The inter-orbital hopping strengths $V_{1,2}$ are depicted as solid green and dashed blue lines, respectively. The inter-site hopping term $H_W$ couples a pair of orbitals of neighbouring sites, depicted as gray lines in Fig. \ref{fig:spectrum}(a). We define Eq.~\eqref{eq:WT} on a coordination-four Mikado lattice, constructed by tracing lines randomly on the plane and placing sites at their intersections (see Appendix \ref{sec:4foldWT}).

Fig.~\ref{fig:spectrum}(a) compares the density of states (DOS) of two inequivalent limits of this model. When $W\gg V_2$ the system is close to the limit of weakly coupled dimmers, which we call the obstructed insulator for reasons that will become clear. It displays mid-gap states around zero energy which are localized at the edges of the system or at the interface with the second type of insulator, realized when $V_2\gg W$, see Fig. \ref{fig:spectrum}(b). This is the atomic insulator limit because the system is close to a set of independent atoms with four orbitals each. 

The mid-gap localized states are reminiscent of those of the SSH model. Indeed, when $V_1\rightarrow 0$, the system reduces to a set of decoupled SSH chains with alternating couplings $V_2$ and $W$. By construction these chains always end on a site, which posses an inner $V_2$ hopping term.
Thus, when $W\gg V_2$, the orbitals couple along the bonds between sites, leaving an isolated orbital at each end of a chain, which form the edge state.
Since amorphous systems are isotropic, this edge state appears all along the boundary (see Fig. \ref{fig:spectrum}(b)).

\begin{figure}
    \centering
    \includegraphics[width = \columnwidth]{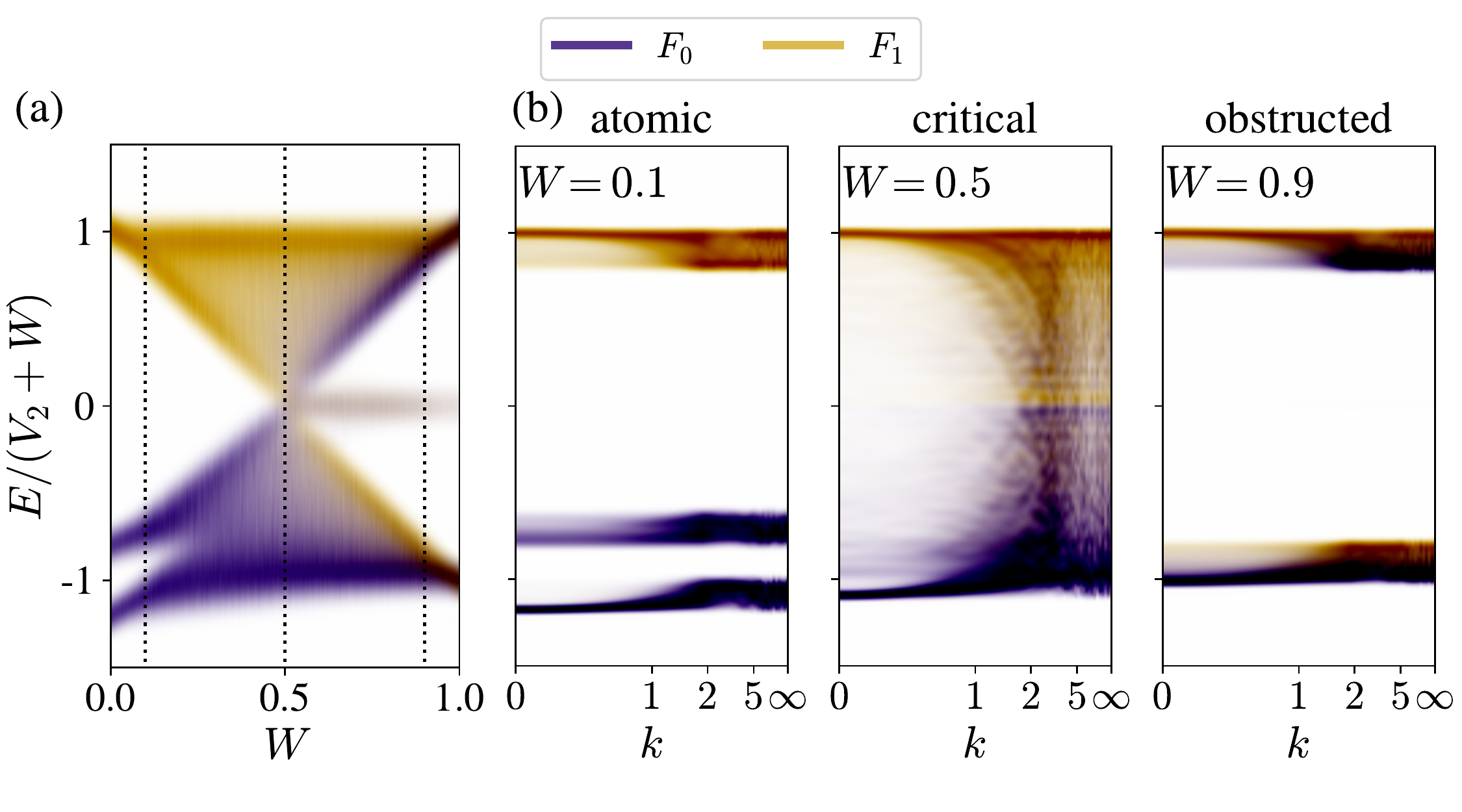}
    \caption{(a) Symmetry projected DOS, $F_{0,1}$. The sub-index $m = 0,1$ indicates the projection to the eigenvalues of the onsite twofold rotation symmetry, $e^{i\pi m}=\pm1$. For $W>0.5$, a localized mid-gap state appears, shown in gray. The vertical dotted lines correspond to the three plots in (b). (b) Momentum-resolved symmetry-projected density of states for the atomic and obstructed limits ($W = 0.1$ and $W = 0.9$, respectively) and the transition between the two ($W = 0.5$). When $W\ll V_{1,2}$ , the occupied bands contain states that project on the symmetric eigenstates of $H_V$. At the transition the two bands cross and mix for $W > 0.5$. The occupied states in this case project onto the symmetric states of $H_W$ (Fig. \ref{fig:bondinv_projection})}
    \label{fig:colouredDOS}
\end{figure}

The two limits are also distinguished by the position of the charge centers of occupied states, $n(\mathbf{r})$, defined in Appendix \ref{app:Fourfold}, and shown in Fig. \ref{fig:spectrum}(c). For $W\gg V_2$, and $W\ll V_2$, the eigenstates show nodes on the bond centers and sites, respectively, which are their symmetry centers with respect to inversion.

The analysis of $n(r)$ suggests a symmetry explanation for the existence of localized edge states. To discuss it, in Fig.~\ref{fig:colouredDOS}(a) we show the symmetry-labelled energy spectrum as a function of $W$~\cite{Marsal2020}.
We color-code it using $F_{0,1}$, which are the projectors onto the eigenstates of $H_V$ with $C_2$-symmetry eigenvalues $\pm1$, introduced in Ref.~\cite{Marsal2020} (see Appendix \ref{sec:4foldWT}).
In Fig. \ref{fig:colouredDOS}(a) we see that when $W\ll V_2$, the eigenstates in the occupied band have a strong overlap with the eigenstates of $H_V$ with symmetry eigenvalue $1$. 
This explains why the charge centers are strongly localized at the atomic sites, as in the left panel of Fig.~\ref{fig:spectrum}(c).
As $W$ increases, the two bands mix: the occupied states project on $H_V$ eigenstates with different symmetry eigenvalues $\pm 1$, while they project on $H_W$ eigenstates with a single symmetry eigenvalue with respect to bond inversion (see Appendix \ref{sec:4foldWT}).
Correspondingly, the right panel in Fig.~\ref{fig:spectrum}(c) shows that the charge centers have shifted to the bond centers.

The crossing of bands is apparent in Fig.~\ref{fig:colouredDOS}(b), where we show the spectrum of the Hamiltonian projected onto plane-waves with well defined momentum $k$ (see Appendix~\ref{app:Heff}). This projection is useful to visualize band inversions between bands labelled by different symmetries~\cite{Marsal2020}. The inversion of the band, occurring at $k\to\infty$, indicates that the two limits are disconnected classes of insulators, and explains why the edge states appear. Hence, although infinitesimal onsite disorder will push these modes away zero energy, their existence is enforced by symmetry, as in crystalline insulators. 

The presence of localized edge states is also signaled by a change in polarization density $P$~\cite{vanderbilt2018berry}, which measures the charge center displacement within a unit-cell in a crystal, or the elemental building block in Fig.~\ref{fig:spectrum}(a) in our amorphous case. For the atomic limit $P=0$, while for the obstructed limit $P = e/2$ (see Appendix \ref{sec:4foldWT}), consistently with the apparition of edge states in the obstructed limit, seen in Fig.~\ref{fig:colouredDOS}(a).

\textit{3D Obstructed phases and surface flat bands in phase-change materials.}
\begin{figure}
    \centering
    \includegraphics[width = \columnwidth]{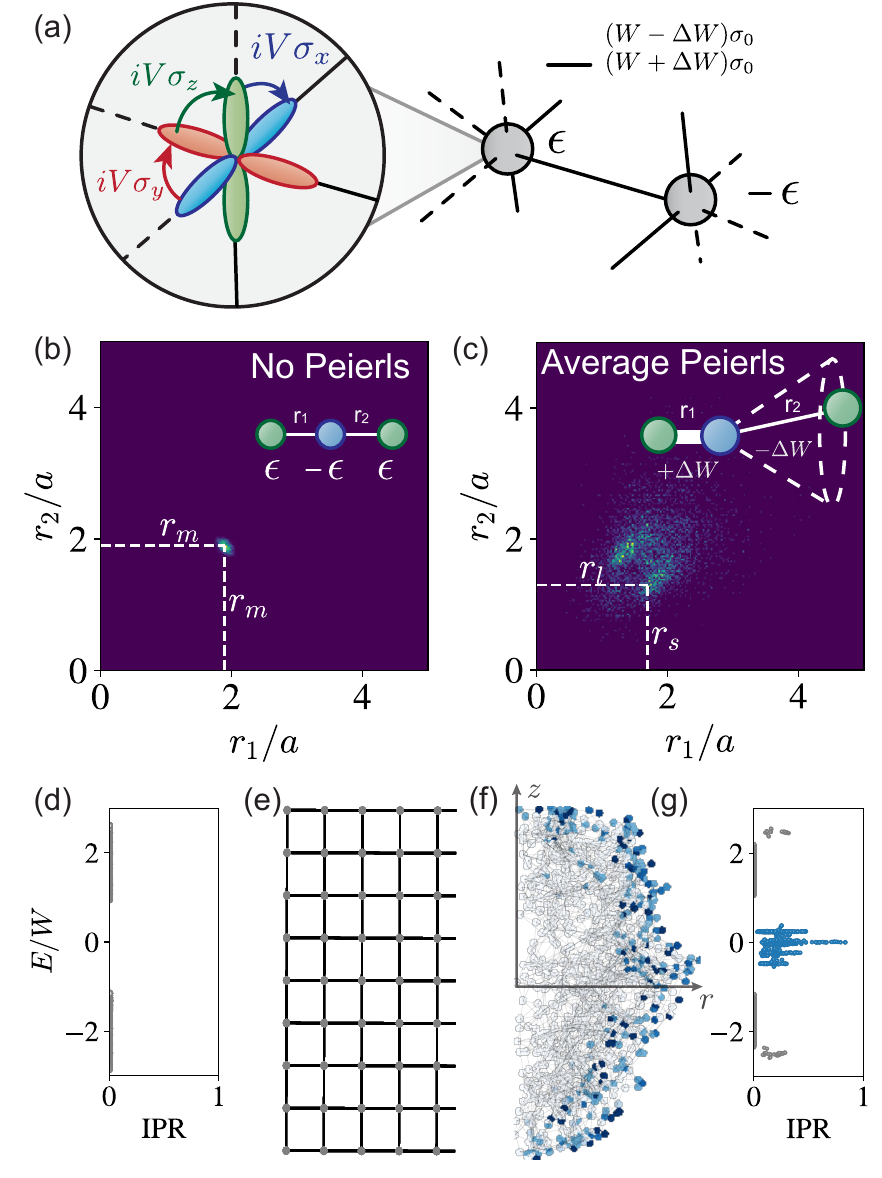}
    \caption{(a) 3D GeTe model. $V$ represents an on-site spin-orbit coupling,  $W\pm\Delta W$ controls the average dimerization and $\epsilon$ is a staggered potential. (b)-(c) Three-particle correlation function $p(r_1,r_2)$, where $(r_1,r_2)$ are pair of distances from a site to opposite neighbours. In the absence of a Peierls distortion there is a single equilibirum distance $r_m$ (b). In the Peierls dimerized phase there are alternating long ($r_l$) and short ($r_s$) equilibrium distances (c). Strong and weak hoppings are associated to shorter and longer bond lengths. (d) and (g) show the inverse participation ratio for the systems in (b) and (c), respectively. (e) and (f) show the local density of in-gap states for 3D system projected into the $(z,r)$ plane. These are absent for the system in (b) and present for the obstructed limit in (c). The system in (b,d,e) has $V = 0.5$, $W = 1$ and $\epsilon = 1.4$, while (c,f,g) has $V = 0.5$, $W\pm\Delta W = 1\pm 1/3$ and $\epsilon = 0$.
    }
    \label{fig:peierls}
\end{figure}
In crystals, one of the mechanism that creates obstructed insulators is a Peierls distortion~\cite{peierls1955quantum}, as in the 1D SSH chain~\cite{Su79}.
It occurs when grouping atoms, chains or layers, is energetically favorable compared to having a single equilibrium distance between them.
However, the Peierls distortion is not exclusive to crystals. In H$_2$ gas~\cite{gaspard2016structure} the hydrogen atoms pair to form molecules, and an average Peierls distortion occurs in amorphous solids~\cite{gaspard2016structure}, including phase-change materials~\cite{Raty2015,muralidharan2022investigation}. Phase-change materials can reversibly switch from crystalline to amorphous states by applying light or current~\cite{leGallo2020overview}. Either or both phases can display a Peierls distortion.
Molecular dynamics simulations show that several phase-change materials, including GeTe, tend to develop a stronger Peierls distortion as they age in the amorphous state~\cite{leGallo2020overview, Raty2015}.

An amorphous Peierls distortion is diagnosed by the three-particle correlation function $p(r_1, r_2)$~\cite{gaspard2016structure}. For a given site, $p(r_1, r_2)$ measures the probability to find a pair of opposite neighbours at distances $r_1$ and $r_2$ respectively, see inset of Fig.~\ref{fig:peierls}(c). 
Two neighbours of a central site are opposed if the angle between their bonds towards the central site is close to $\pi$.
If the system is not distorted, $p(r_1, r_2)$ peaks at a single nearest-neighbour average distance $r_m$, as shown in Fig. \ref{fig:peierls}(b) for a perfectly crystalline lattice. However, if an average Peierls dimerization occurs, $p(r_1, r_2)$ peaks at two different average values, $r_s$ and $r_l$, corresponding to shorter and longer bonds, respectively (Fig.  \ref{fig:peierls} (c)).

We now introduce a tight-binding model suitable to describe phase-change materials that crystallize in a cubic rock-salt-like structure and display an average Peierls distortion in the amorphous phase. This is the case of GeTe~\cite{Lent1986} or Ge$_2$Sb$_2$Te$_5$~\cite{muralidharan2022investigation}. In this model, described in detail in Appendix \ref{app:3DmodelGeTE}, each atom has six neighbours, sketched in Fig. \ref{fig:peierls}(a). The amorphous lattice is constructed by a 3D generalization of the random-line method: sites are at the intersection of three planes with random orientations. 
At each site there are three $p$ orbitals, coupled by spin-orbit coupling through $H_V = V \mathbf{L\cdot S}$, and aligned with the direction of their neighbours.
Nearest neighbours are coupled through a $\sigma$-hopping between the two orbitals aligned with the bond direction. 
The hopping strength is $W\pm \Delta W$ for weak and strong hoppings, alternating along approximately aligned bonds. The on-site staggered potential is $\pm \epsilon$, such that nearest neighbours have opposite onsite energy.

\begin{figure}
    \centering
    \includegraphics[width = \columnwidth]{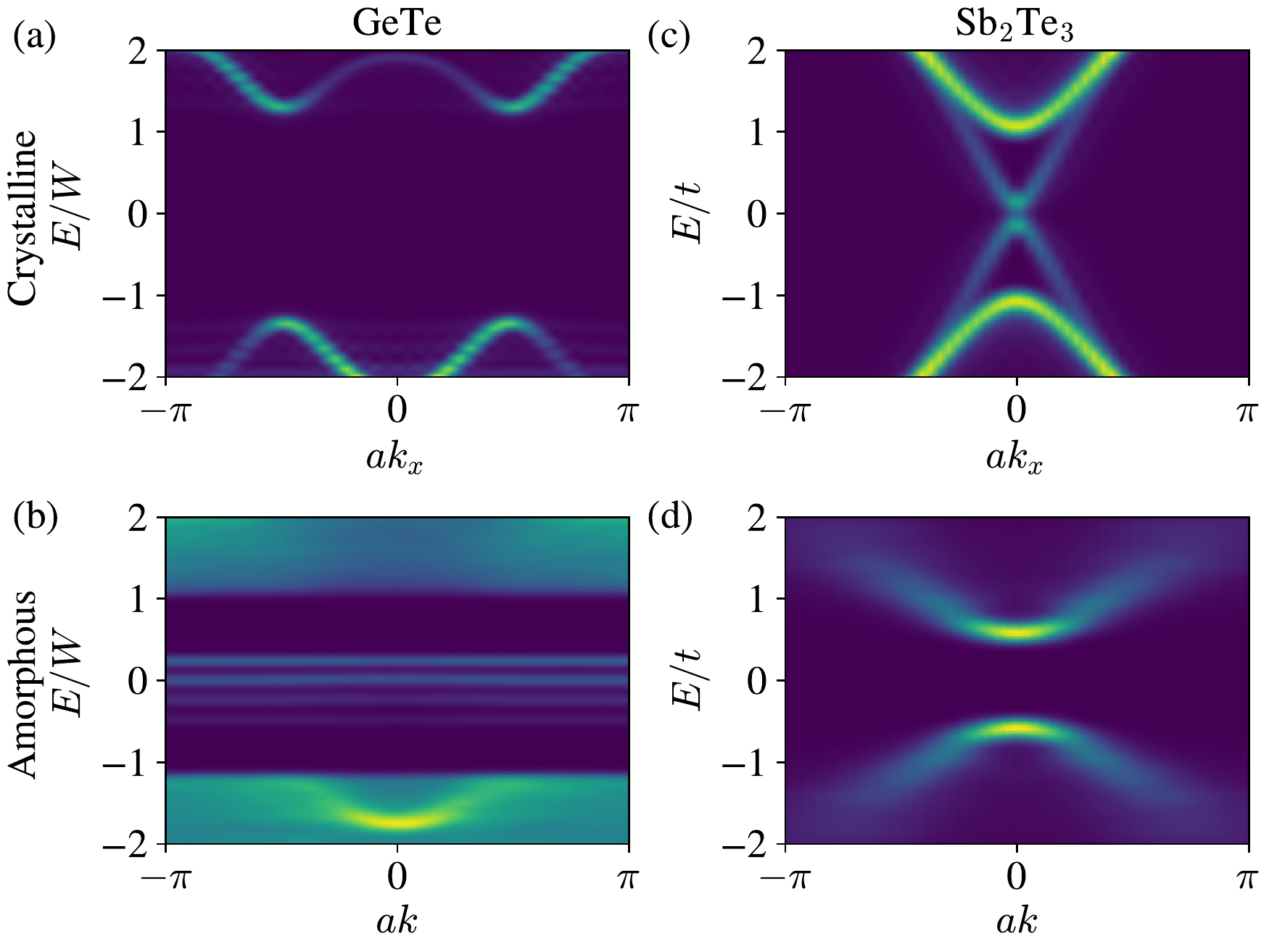}
    \caption{Spectral function for phase-change materials in the crystalline undistorted (a-c) and amorphous distorted (b-d) phases. $a$ is the cubic lattice spacing in the crystalline systems and the average nearest-neighbour distance in the amorphous samples. In the model of GeTe (a-b), the distortion closes and reopens the gap, resulting in a localized surface state. The parameters for the two plots are the same as in Fig. \ref{fig:peierls}. In the model of Sb$_2$Te$_3$ (c-d), the distortion gaps out the topological surface state. The parameters for this system are $M = 5$ and $\Delta t = 0.8$ (see App. \ref{app:3DBHZ} for details).}
    \label{fig:ARPES}
\end{figure}

Assuming, as in GeTe, that there is no Peierls distortion in the crystal ($\Delta W=0$), the spectrum has two bands at  $\epsilon\neq0$ (see Fig.~\ref{fig:peierls}(d)).
This parameter choice breaks inversion symmetry with respect to bond centers but preserves it with respect to the sites. By taking $\epsilon/W\rightarrow \infty$ we smoothly localize the charge on atomic sites, realizing the trivial atomic limit. Accordingly, we find no surface states (see Fig.~\ref{fig:peierls}(d) and (e))

With a finite distortion ($\Delta W\neq0$) but no staggered potential ($\epsilon = 0$), the charge center of the occupied bands shifts toward the bond centers. In a finite system, the surface intersects those bonds. Hence the atoms form dimmers and, as in our 2D model, the occupied states have a non-zero polarization $P = e/2$. We thus expect localized surface states around half-filling, which are seen in Fig.~\ref{fig:peierls} (f) and (g).

The inverse participation ratio (IPR) $p(E) = \sum_i\sum_j \left|\Psi_{ij}(E)\right|^4$
is a measure of the localization of the edge states. The mid-gap states in the obstructed limit have a much higher IPR than the bulk band states (see Fig.~\ref{fig:peierls}(g)), indicating that they involve fewer sites.
This is confirmed by the local DOS, which shows that the sites involved in the in-gap localized states are localized at the surface of the system (see Fig.~\ref{fig:peierls}(f)). In contrast, the crystal bulk states are not localized, and no surface states appear (see Fig.~\ref{fig:peierls}(d) and (e)). In Appendix \ref{app:3DmodelGeTE} we show how the obstructed and atomic insulator limits are also symmetry indicated, as for the 2D case.

While the mechanism that induces localized in-gap states is similar to that of a crystal, the difference lies in which surface they appear. For a crystal, surface states appear at facets that respect the crystalline (Peierls) symmetry. In the 3D amorphous obstructed state, this symmetry is satisfied on average~\cite{Spring2021}, and thus in-gap localized states appear no matter how the sample is terminated, as we observe in Fig.~\ref{fig:peierls}(f). This observation suggests that a more spectacular signature can manifest in their surface spectral function, measurable in angle-resolved photoemission experiments (ARPES) even in amorphous materials~\cite{Morgan1983,Corbae:2019tg}. Fig. \ref{fig:ARPES} shows the spectral function, calculated by projecting the real-space spectral function onto a plane wave basis: $A(\mathbf{k},E) = -\frac{1}{\pi}\mathrm{Im}\bra{\mathbf{k}}(\hat{H}-E)^{-1}\ket{\mathbf{k}}$.
The states $\ket{\mathbf{k}}$ are plane waves defined with a phase $e^{i\mathbf{k\cdot r_i}}$ on site $i$, and the momentum $\mathbf{k}$ represents the momentum of the photo-emitted electron. 

Fig.~\ref{fig:ARPES} (a) and (b) compare the spectral function for crystalline and amorphous GeTe models. The crystal is fully gaped due to a finite staggered potential. The amorphous phase is dominated by the Peirels distortion, resulting in two-dimensional surface flat bands around zero energy. The surface flat bands merge into a single flat band as $V \to 0$, indicating that the gaps between them originate from the different ways of terminating the sample. The surface flat bands appear at the energies where the localized surface states are seen in Fig.~\ref{fig:peierls}(f) and (g).  Their emergence can be thus controlled by transitioning from crystal to amorphous.

Not all Peierls distortions generate obstructed limits. In Fig.~\ref{fig:ARPES}(c) and (d) we compare the spectral function for crystalline and amorphous Sb$_2$Te$_3$, one of the first predicted 3D topological insulator crystals~\cite{Zhang:2009ks,Zhu2015}, which also is a phase-change material~\cite{Guo:2018bf,Korzhovska2020}. The crystal displays the characteristic surface Dirac cone. When the Peierls distortion dominates in the amorphous phase, the spectrum is fully gaped, as in the atomic limit. Unlike for our GeTe model, the Peierls phase of Sb$_2$Te$_3$ is not smoothly connected to a set of decoupled SSH chains. This phenomenology is consistent with recent experiments suggesting that the topological properties are lost when Sb$_2$Te$_3$ phase-changes into the amorphous form~\cite{Korzhovska2020}.

\textit{Discussion.} The models and tools we described showcase amorphous obstructed limits, and the controllable emergence of surface flat bands in phase-change materials. They are starting points to develop material-specific models, that may include effects that we neglected, such as fluctuations in the local coordination of sites~\cite{Raty2015}, or the well-understood effects of non-geometric disorder~\cite{Wu_2016}, including spatial variations in the hopping or on-site potential.

In summary, we have defined amorphous obstructed insulators as a necessary ingredient to understand phase-change materials. Our results establish that this class of materials represents a promising platform to switch on and off topological properties, and two-dimensional surface flat bands. \\

\textit{Acknowledgements.} 
We are grateful to S. Ciocys, P. Corbae, D. Mu\~{n}oz-Segovia, R. Queiroz and S. Tchoumakov for discussions.
A.G.G. acknowledges financial support from the European Union Horizon 2020 research and innovation program under grant agreement No. 829044 (SCHINES).
D.V. acknowledges funding from the Swedish Research Council (VR) and the Knut and Alice Wallenberg Foundation. \\ 

\textit{Author contributions, and data and code availability.} 
Q.M. performed the calculations assisted by D.V. and A.G.G..
Q. M. wrote the manuscript with input from all authors. 
A.G.G. devised and supervised the project. The data and codes that support the findings presented in this work are available from~\cite{zenodo}.

%

\clearpage
\newpage

\setcounter{secnumdepth}{5}
\renewcommand{\theparagraph}{\bf \thesubsubsection.\arabic{paragraph}}

\renewcommand{\thefigure}{S\arabic{figure}}
\setcounter{figure}{0} 

\appendix

\onecolumngrid
\section{\label{app:Fourfold} Fourfold coordinated Weaire-Thorpe lattice}
\label{sec:4foldWT}

The fourfold-coordinated Weaire-Thorpe system is built similarly to that discussed in the Supplementary information of Ref.~\cite{Marsal2020}. We first build the lattice using the method described in \cite{Miles901}. This algorithm strictly enforces the constant coordination property. Then we set the Hamiltonian onto it. 
The Hamiltonian coefficients do not depend on the position of the sites nor on the orientation of the bonds they involve. 
Thus, as in all Weaire-Thorpe amorphous systems, disorder is only structural.

\subsection{Lattice}
The lattice is drawn from a mikado of $N$ random lines, $N$ being chosen according to a Poisson distribution with mean $2R\sqrt{\pi \rho}$. 
$\rho$ is the average density of lines in the system, and $R$ is its radius.
Each line has a random slope $\theta \in \left[0,2\pi\right]$ and a random y-intercept.
The intersection of pairs of lines become the sites of the lattice and the links between them are given by the segments of lines. 
The constant coordination of the lattice is guaranteed by the fact that having a coordination $2n$ requires that $n$ lines cross at the same point, which is unlikely for $n>2$ in a random lattice.

\subsection{Hamiltonian}

The Hamiltonian of the fourfold-coordinated lattice is a Weaire-Thorpe Hamiltonian \cite{Weaire1970}. It decomposes into an onsite term $H_V$ and a constant hopping term $H_W$:
\begin{equation}
    H = \sum_i \mathbf{c}_i^\dagger H_V \mathbf{c}_i + \sum_{\left<i,j\right>}\mathbf{c}_i^{\dag} H_W(i,j)\mathbf{c}_j,
\end{equation}
Each site has four orbitals that overlap with each other according to $H_V$. A single pair of orbitals belonging to neighbouring sites is also coupled through the $H_W$ term. $\mathbf{c}_i$ is the vector representing the four orbitals of site $i$. The two terms of the Hamiltonian read:
\begin{equation}
    H_V = \begin{pmatrix} 0&-V_1&-V_2&-V_1\\-V_1&0&-V_1&-V_2\\-V_2&-V_1&0&-V_1\\-V_1&-V_2&-V_1&0\end{pmatrix}, \quad
    H_W(i,j) = -W\epsilon_{ij}.
\end{equation}

$\epsilon_{ij}$ is a $4\times 4$ matrix whose coefficients are all zero except the two that indicate which orbitals of sites $i$ and $j$ link the two sites.

\begin{figure}
    \centering
    \includegraphics[width = \textwidth]{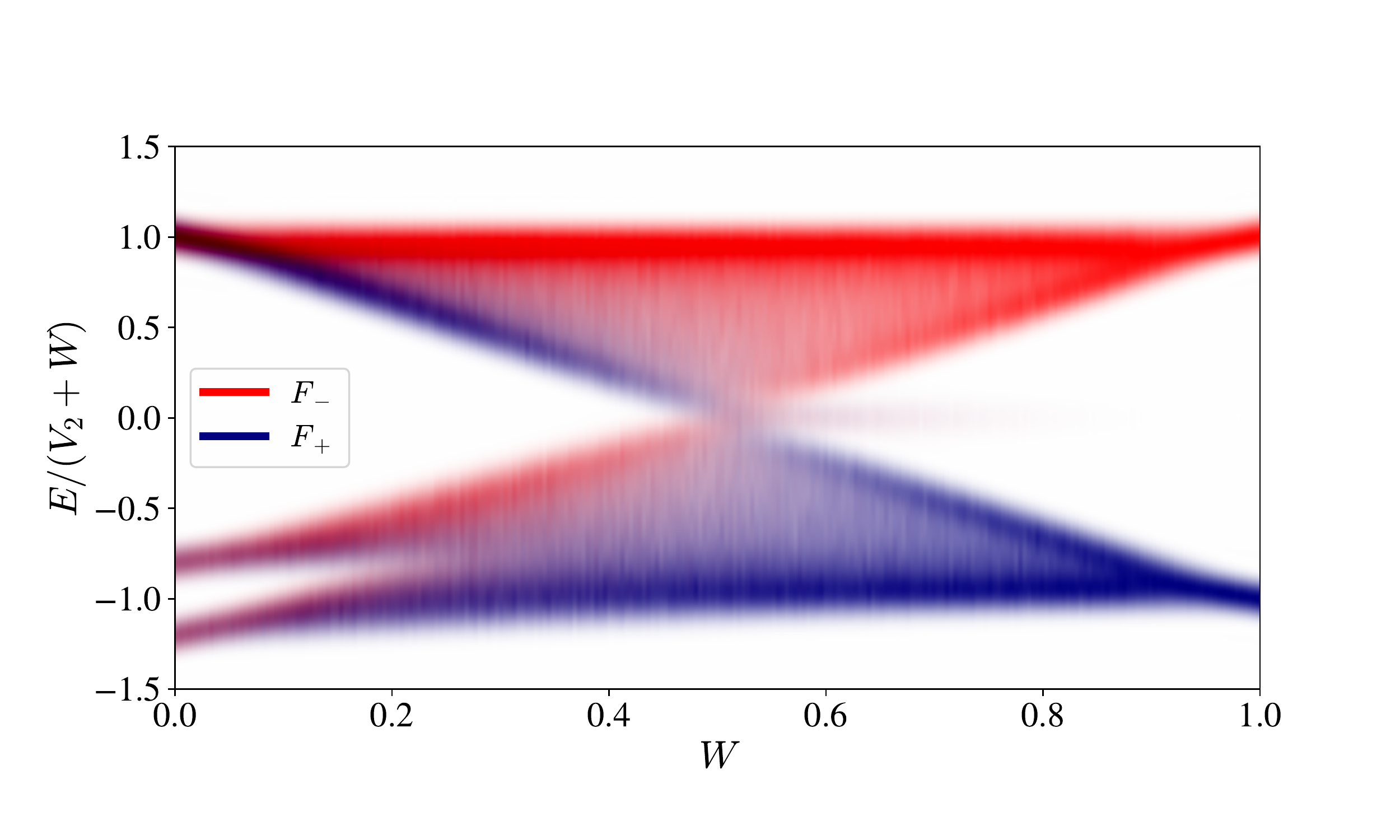}
    \caption{Density of states for the 2D system (Hamiltonian \eqref{eq:WT}) projected on the symmetric ($F_+$) and anti-symmetric ($F_-$) eigenstates of each bonds. Similarly to Fig. \ref{fig:colouredDOS}, when $W\gg V_{1,2}$ the spectrum is organized in two bands whose states project on a single eigenstate of each bond. The projection of the edge states vanishes since the edge sites have no neighbours to define symmetric and anti-symmetric eigenstates.}
    \label{fig:bondinv_projection}
\end{figure}

This system has two trivial limits. 
When $W = 0$, it is a set of independent sites, with four obitals each. The eigenstates of $H_V$ are $(1,i^m, -1^m, -i^m)^T$ for $m = 0,1,2,3$ and their energy is $-V_2+2V_1,\ V_2,\ -V_2-2V_1,\ V_2$.
We define $F_{0,1}$ such that:
\begin{eqnarray}
F_{0}(\ket{\psi}) =& \sum_{\mathrm{sites}\ i}\left|\braket{i,0|\psi}\right|^2+\left|\braket{i,2|\psi}\right|^2 = \braket{\psi|\hat{F_0}|\psi},\\
F_{1}(\ket{\psi}) =& \sum_{\mathrm{sites}\ i}\left|\braket{i,1|\psi}\right|^2+\left|\braket{i,3|\psi}\right|^2= \braket{\psi|\hat{F_1}|\psi},
\end{eqnarray}
where $\ket{i,m}$ is the $m^{\mathrm{th}}$ eigenstate of $H_V$ localized on the site $i$.
The density plot on Fig.~\ref{fig:colouredDOS} is then given by $F_{0,1}(E) = \mathrm{Tr}(\delta(\hat{H}-E)\hat{F}_{0,1})/N$.
The other trivial limit occurs when $V_{1,2}=0$: the system decomposes in the set of decoupled dimers with coupling strength $W$ between the two orbitals.
One can analyze the symmetry of the system by projecting the density of states onto the local eigenstates of $H_W$:
\begin{eqnarray}
    F_{+}(\ket{\psi}) =& \sum_{\mathrm{bonds}\ j}\left|\braket{j,+|\psi}\right|^2 = \braket{\psi|\hat{F_+}|\psi},\\
    F_{-}(\ket{\psi}) =& \sum_{\mathrm{bonds}\ j}\left|\braket{j,-|\psi}\right|^2 = \braket{\psi|\hat{F_-}|\psi},
\end{eqnarray}
where $\ket{j,\pm}$ is the eigenstate of $H_W$ localized on the bond $j$ with eigenvalues $\mp W$.
$\ket{j,+}$ is a the eigentate of $H_W$ that is symmetric with respect to bond inversion while $\ket{j,-}$ is anti-symmetric.
$F_+$ and $F_-$ are shown in Fig.~\ref{fig:bondinv_projection}.

\subsection{Charge centers}

The charge center in Fig. \ref{fig:spectrum}(c) of the main text are calculated as
\begin{equation}
    n(\mathbf{r}) = \sum_m |\sum_i\sum_j\Psi^m_{ij} a_{ij}(\mathbf{r})|^2/N_{\mathrm{occ}},
\end{equation}
where $m$ runs over the $N_{\mathrm{occ}}$ eigenstates $\Psi^m$. The function
\begin{equation}
    a_{ij}(\mathbf{r}) = \exp\left(-\frac{(\mathbf{r-r}_i)^2+\frac{3}{2}(\mathbf{r-r}_{ij})^2}{2(\mathbf{r}_{ij}-\mathbf{r}_i)^2}\right),
\end{equation}
encodes the spatial dependence of the local orbital $j$ on site $i$ as a function of $\mathbf{r}_i$, the position of site $i$, and $\mathbf{r}_{ij}$, the center of the bond involving the orbital $j$ of $i$.

\subsection{Polarization}

The presence of localized edge states can also be understood by focusing on the polarization, which measures the average position of charges compared to the unit cell center~\cite{vanderbilt2018berry}. In inversion symmetric systems, charges necessarily lie at inversion symmetric points. If they are localized on the unit cell center, the polarization density $P$ is an integer in units of the electron charge $e$. If, on the contrary, they are localized at the unit cell boundaries, $P$ is a half-integer.  

The value of $P$ is relative to the choice of unit cell, which can be centered either on the inner hopping within sites ($V$-bonds) or on the outer hopping between sites ($W$-bonds).
To simplify the analysis, it is convenient to set the unit cell consistently with the termination of the system as discussed in Refs.~\cite{deJuan2014, Ryu02}. 
In our system, the edge always falls on a bond between sites, cutting a $W$-bond.
We then choose the building block centered on sites, as shown in the panel of Fig.~\ref{fig:spectrum}(a).
Following \cite{deJuan2014}, the number of mid-gap edge states $n$ is given by $n\equiv 2P \ \mathrm{mod}\ 2$. Thus, from Fig.~\ref{fig:spectrum} (c) we see that $P=0$ for the atomic limit and $P = e/2$ for the obstructed one. This analysis is consistent with the apparition of edge states in the obstructed limit, seen in Fig.~\ref{fig:colouredDOS}(a).

\section{\label{app:3DmodelGeTE}Local symmetries of the GeTe Weaire-Thorpe model}

The 3D sixfold-coordinated model is built using the same method as the 2D fourfold-coordinated.
Rather than using a random set of lines, we used a random set of planes. 
Three independent planes necessarily cross at a single point, through which three lines pass, defined by the intersections of two of those planes.
Thus, the sites of the system will be the intersections between three planes, and the links between them the intersections between pairs of planes.

\subsection{Hamiltonian}

Each site has three orbitals $p_x$, $p_y$, $p_z$ that can host spin up and spin down electrons. 
Thus, the onsite term of the hamiltonian is a $6\times 6$-matrix which reads:
\begin{equation}
    H_V = \frac{V}{2}\begin{pmatrix}
    0&-i&1&0&0&0\\
    i&0&-i&0&0&0\\
    1&i&0&0&0&0\\
    0&0&0&0&i&-1\\
    0&0&0&-i&0&-i\\
    0&0&0&-1&i&0 
    \end{pmatrix}
    = V {\bm L}\cdot {\bm S} = iV ({\bm c}^\dagger\wedge {\bm c})\cdot {\bm S},
\end{equation}
where we ordered the basis states as $\ket{p_x, \uparrow}, \ket{p_y, \uparrow}, \ket{p_z, \downarrow}, \ket{p_z, \uparrow}, \ket{p_y, \downarrow}, \ket{p_x, \downarrow}$.
On each site, the orbitals $p_{x,y,z}$ are supposed to be aligned  with the bonds towards neighbouring sites. 
Using the same approximation as in Weaire-Thorpe systems, we admit that the hopping terms between orbitals do not depend on the exact orientation of the bonds and orbitals. 
Thus, the hopping strength does not depend on the considered bond, and disorder is only present in the structure of the lattice.
Bonds between sites are therefore made by two $p$ orbitals of neighbouring sites, overlapping with each other to form a $\sigma$ covalent bond. 
Since the labeling $x,y,z$ of the orbitals is local, a $p_x$ orbital of a site $i$ can be bond to the $p_y$ orbital of a neighbour $j$. In this case, the hopping term is of the form 
\begin{equation}
    H_W(i,j) = -(W\pm \Delta W)\begin{pmatrix}
    0&1&0&0&0&0\\
    0&0&0&0&0&0\\
    0&0&0&0&0&0\\
    0&0&0&0&0&0\\
    0&0&0&0&0&1\\
    0&0&0&0&0&0
    \end{pmatrix},
    \label{eq:hopamorph1}
\end{equation}
where the coupling strength $W\pm \Delta W$ depends wether this bond is weak or strong and is determined in a second step.
The two non-zero terms in $H_W(i,j)$ account for spin up and down.
The same orbital $p_x$ of $i$ will also be bound to the opposite neighbour $j'$ of $i$, defined so that $j$, $i$ and $j'$ are approximately aligned in this order.
Again, since the labeling $x,y,z$ of the orbitals is local, it can bound to any of the three orbitals of $j'$.
If the bond involves for example the orbital $p_z$ of $j'$, the hopping term reads:
\begin{equation}
    H_W(i,j') = -(W\mp \Delta W)\begin{pmatrix}
    0&0&1&0&0&0\\
    0&0&0&0&0&0\\
    0&0&0&0&0&0\\
    0&0&0&0&0&1\\
    0&0&0&0&0&0\\
    0&0&0&0&0&0
    \end{pmatrix}.
    \label{eq:hopamorph2}
\end{equation}

The sign of $\Delta W$ for this latter bond is necessarily the opposite of that of the former since weak and strong bonds alternate along aligned sites. 
The fact that we can define aligned bond in the amorphous structure comes from the enforced coordination six of the lattice with sites having an octahedral structure.
The algorithm used to determine the strength $\pm \Delta W$ of the bonds can be found in~\cite{zenodo}.

\subsection{Crystalline version}

When set on a cubic lattice, the Hamiltonian of the system can be written with a two-site unit cell \cite{Lent1986, Hsieh:2012tq}:
\begin{equation}
    H = \sum_{i} V (\mathbf{c}_i^\dagger\wedge \mathbf{c}_i)\cdot \mathbf{S} + (-1)^i\epsilon \mathbf{c}_i^\dagger\cdot\mathbf{c}_i - \sum_{<i,j>} (W+(-1)^i\mathrm{sgn}(\hat{d}_{ij}\cdot(\hat{e}_x+\hat{e}_y+\hat{e}_z)\Delta W) (\mathbf{c}^\dagger_i\cdot \hat{d}_{ij})(\mathbf{c}_j\cdot \hat{d}_{ji}),
    \label{eq:ham_aGeTe}
\end{equation}

where $i,j$ labels the sites so that nearest neighbours, denoted by $<\cdots>$, have $(-1)^{i+j} = -1$, and $\hat{d}_{ij}$ is the unit vector pointing from $i$ to $j$. 

The corresponding Bloch hamiltonian is a $12\times12$ matrix (two sites per unit cell, three orbitals per site, with spin $\frac{1}{2}$). It reads
\begin{equation}
    \mathcal{H}(k) = 
    \begin{pmatrix}
        H_{Ge}&H_{hopping}\\
        H_{hopping}^\dagger&H_{Te}
    \end{pmatrix}.
    \label{eq:ham_cGeTe}
\end{equation}
The matrices $H_{Ge}$ and $H_{Te}$ are the same up to the diagonal staggered potential $\epsilon$, which takes opposite values for the two elements. They read:
\begin{equation}
H_{Ge} = \begin{pmatrix}
    \epsilon&-\frac{iV}{2}&\frac{V}{2}&0&0&0\\
    \frac{iV}{2}&\epsilon&-\frac{iV}{2}&0&0&0\\
    \frac{V}{2}&\frac{iV}{2}&\epsilon&0&0&0\\
    0&0&0&\epsilon&\frac{iV}{2}&-\frac{V}{2}\\
    0&0&0&-\frac{iV}{2}&\epsilon&-\frac{iV}{2}\\
    0&0&0&-\frac{V}{2}&\frac{iV}{2}&\epsilon
\end{pmatrix}, \quad 
H_{Te} = \begin{pmatrix}
    -\epsilon&-\frac{iV}{2}&\frac{V}{2}&0&0&0\\
    \frac{iV}{2}&-\epsilon&-\frac{iV}{2}&0&0&0\\
    \frac{V}{2}&\frac{iV}{2}&-\epsilon&0&0&0\\
    0&0&0&-\epsilon&\frac{iV}{2}&-\frac{V}{2}\\
    0&0&0&-\frac{iV}{2}&-\epsilon&-\frac{iV}{2}\\
    0&0&0&-\frac{V}{2}&\frac{iV}{2}&-\epsilon
\end{pmatrix}.
\end{equation}

$H_{hopping}$ couples Germanium and Tellurium sites with alternatively weak and strong bond whose strength is $W\pm \Delta W$. $p$-orbitals only couple along their symmetry axis, hence the hopping term reads:
\begin{equation}
    H_{hopping} = 2W\begin{pmatrix}
        \cos k_x&0&0&0&0&0\\
        0&\cos k_y&0&0&0&0\\
        0&0&\cos k_z&0&0&0\\
        0&0&0&\cos k_z&0&0\\
        0&0&0&0&\cos k_y&0\\
        0&0&0&0&0&\cos k_x
    \end{pmatrix}
    +2i\Delta W \begin{pmatrix}
        \sin k_x&0&0&0&0&0\\
        0&\sin k_y&0&0&0&0\\
        0&0&\sin k_z&0&0&0\\
        0&0&0&\sin k_z&0&0\\
        0&0&0&0&\sin k_y&0\\
        0&0&0&0&0&\sin k_x
    \end{pmatrix}.
\end{equation}

In the crystalline case, it is possible to define $x$, $y$, and $z$ directions consistently with the orientation of the $p$-orbitals in the whole lattice, hence the simple form of the hopping term. 
This is not possible in the amorphous system, where the $p_x$ orbital of a site can be aligned with the $p_y$ orbital of its neighbour, leading to off-diagonal terms in the hopping part of the Hamiltonian (Eqs. \ref{eq:hopamorph1}, \ref{eq:hopamorph2}).

When $\epsilon \neq 0$, this model is gapped at half-filling, as Fig.~\ref{fig:3Ddos}b.
If $\Delta W$ is non zero, {\it i.e.} when the distortion occurs, localized surface states arise at zero energy. 
Those states are pinned at zero energy if $\epsilon$ is zero. 
Otherwise, they are gapped out and merge with the bulk bands as $\epsilon$ increases.

\subsection{Atomic and obstructed limits}

\begin{figure}
    \centering
    \includegraphics[width = \columnwidth]{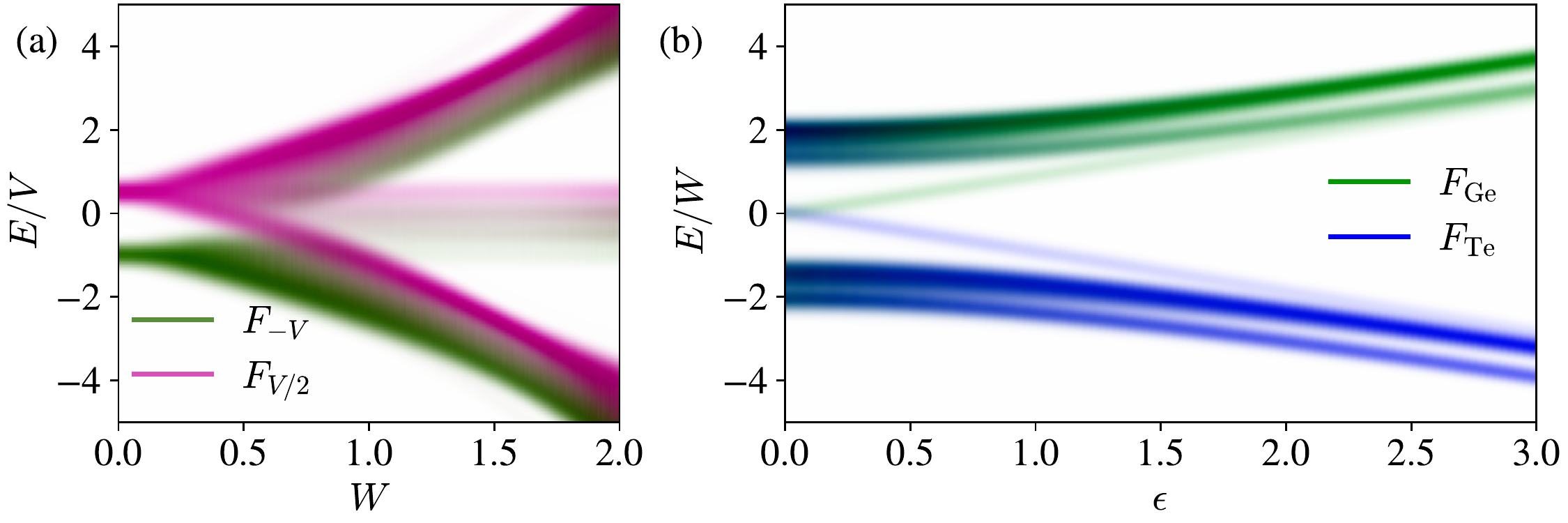}
    \caption{(a) Symmetry projected density of states for the amorphous Hamiltonian \eqref{eq:ham_aGeTe}. $F_{-V}$, $F_{V/2}$ are the projection on the eigenstates of the isolated atoms with energy $-V$, $V/2$. When the hoppings are set to zero, the system has two degenerate bands. When $W/V$ and $\Delta W/V$ increase, the gap closes, bands mix and mid-gap states appear. Here, we set $\Delta W = 2W/3$. (b) Density of states for the crystal GeTe model set on a cubic lattice (Hamiltonian \eqref{eq:ham_cGeTe}) as a function of $\epsilon$, for $V = 0.5W$ and $\Delta W = 2W/3$. The green and blue colors show the weight of each sublattice. The degeneracy of the surface states is lifted when the onsite staggered potential $\epsilon$ is set on. The surface states then merge with the bulk bands for high $\epsilon$.} 
    \label{fig:3Ddos}
\end{figure}

The onsite term $H_V$ has six eigenstates that form two degenerate bands with energies $\frac{V}{2}$ and $-V$. 
Fig.~\ref{fig:3Ddos}a shows the spectrum of the system and the projection of its eigenstates on that of $H_V$ for $W$ ranging from 0 to 3 and $\Delta W = W/3$. 
If $\ket{i,m}$ are the eigenstates of $H_V$ localized on site $i$ with energy $-V$ if $m=0,1$ and $V/2$ if $m = 2,...,5$, then:
\begin{eqnarray}
    F_{-V}(E) = \sum_i\sum_{m = 0}^1 |\braket{\psi(E)|i,m}|^2,\\
    F_{V/2}(E) = \sum_i\sum_{m = 2}^5 |\braket{\psi(E)|i,m}|^2.
\end{eqnarray}
For small enough $W$, the system has two bands corresponding to that of the atomic limit.
When $W \pm \Delta W$ increases, the gap closes and then reopens, with localized mid-gap states appearing. 
The projectors on local eigenstates of $H_V$ show that the two atomic bands have mixed to form an obstructed limit.

Figure~\ref{fig:3Ddos}b shows how the energy bands evolve once the staggered potential is turned on. It shows the density of states, coloured with the relative weight of each sublattice, defined by the sign of the potential. For low $\epsilon$, the surface state is split in two before merging into the two bands. When $\epsilon$ increases further, the bulk states of each band are localized on only one of the two sublattices. 
This is a second atomic limit for the system, on which we focus in the crystalline model of GeTe (Fig.~\ref{fig:peierls}).

\section{\label{app:3DBHZ} Bernevig-Hughes-Zhang model for Sb$_2$Te$_3$}

The model for Sb$_2$Te$_3$ is built using the same lattice as GeTe. It is therefore a sixfold coordinated amorphous lattice, now with four orbitals per site. We use the same algorithm as in the previous case to define weak and strong bonds. The onsite term of the Hamiltonian reads:
\begin{equation}
    \hat{H}_{os} =
    M\sigma_0 \otimes \tau_z,
    \end{equation}
and the hopping amplitude between sites $i$ and $j$ is 
\begin{equation}
    H(\hat{d}_{ij}) = (t\pm \Delta t) (\sigma_0\otimes\tau_z+i\hat{d}_{ij}\cdot\overrightarrow{\sigma}\otimes\tau_x).
\end{equation}

This model has two main features that are different from that of GeTe.
First of all, it is not strictly a Weaire-Thorpe type Hamiltonian because, here, the hopping term does depend explicitely on the bond direction $\hat{d}_{ij}$. 
Thus positional disorder adds to structural disorder to make the system amorphous.
Second, this system is not equivalent to a set of independent SSH chains when the onsite part is driven to zero. 
The four orbitals of each site are all involved in the bonds towards the six neighbours.
This is why, when the distortion is on ($\Delta t \neq 0$), the topological surface state gaps out, but no localized flat band appear within the gap, as seen in Fig. \ref{fig:ARPES} of the main text.

If it was set on a crystalline, cubic lattice, the Bloch Hamiltonian of the system would be
\begin{equation}
    \mathcal{H}(\mathbf{k}) = \begin{pmatrix}
        H_0&H_{hopping}\\
        H_{hopping}^\dagger&H_0
    \end{pmatrix},
\end{equation}
with
\begin{eqnarray}
    H_0 &=& M\sigma_0\otimes\tau_z\\
    \nonumber
    H_{hopping} &=& 2t \left(\sum_i \cos k_i \sigma_0\otimes\tau_z-\sum_i \sin k_i \sigma_i\otimes\tau_x \right)\\
    &+& 2i\Delta t \left(\sum_i \sin k_i \sigma_0\otimes \tau_z + \sum_i \cos k_i \sigma_i\otimes \tau_x\right).
\end{eqnarray}
When $\Delta t = 0$, the original Bernevig-Hughes-Zhang model is recovered. Depending on the relative strengths of the BHZ model parameters compared to the Peierls distortion different limits are possible in the amorphous and crystalline phases: a trivial atomic insulator (large staggered potential $M$), an trivial insulator (large Peierls distortion $\Delta t$), and topological phases for intermediate parameters. 

\section{Effective Hamiltonian and momentum-resolved symmetry-projected spectral functions}
\label{app:Heff}
We can understand our findings in terms of the effective Hamiltonian, which is an efficient tool to analyze symmetries and diagnose topology in disordered materials~\cite{Varjas2019, Marsal2020}.
We define the effective Hamiltonian $H_{\rm eff}(\bk) = G_{\rm eff}(\bk)^{-1} + E_F$ through the projection of the single-particle Green's function onto plane-wave states
\begin{equation}
\label{eq:Geff}
G_{\rm eff} (\bk)_{l,l'}  = \bra{\bk, l} G \ket{\bk, l'},
\end{equation}
where $G = \lim_{\eta\to 0} \left( H - E_F + i \eta\right)^{-1}$ is the Green's function of the full Hamiltonian with $E_F$ chosen to be in a gap.
The states $\ket{\bk, m}$ are normalized plane-wave states with on-site angular momentum indexed by $m$ (in the 2D WT model this index corresponds to representations of fourfold rotation, while in the 3D case to a combination of the total angular momentum and angular momentum $z$ component quantum numbers of a spinful $p$-orbital). 
In the real space basis these states are given by
\begin{equation}
\label{eq:plane_wave}
\ket{\bk, m} = \sum_i \frac{1}{\sqrt{N}}\exp(i \bk \br_i) \ket{i, m},
\end{equation}
where $\br_i$ is the position of site $i$, and $N$ is the number of sites in the sample.
States with different $m$ are orthogonal. The basis is, however, overcomplete with respect to $\bk$, because the typical overlap between different $\bk$ states with the same $m$ only decays as $1/\sqrt{N}$ when approaching the thermodynamic limit.

A central property of $H_{\rm eff}$ is that its gap closes only when the gap of the full Hamiltonian closes.
This follows from the fact that $H_{\rm eff} - E_F$ can only have a zero if $G_{\rm eff}$ has a pole, which is only possible if $G$ has a pole, when $H - E_F$ has a zero.
Hence, a topological invariant defined in terms of $H_{\rm eff}$ that can only change when its gap closes is also a good topological invariant for the original system~\cite{Varjas2019, Marsal2020}.
In the large $|\bk|\equiv k$ limit the expectation value in \eqref{eq:Geff} reduces to purely on-site terms, as the relative phases between different sites average to $0$ in the thermodynamic limit, resulting in $H_{\rm eff}(k=\infty)$ being identical to $H_{\rm eff}^{W=0}(k=0)$ in the system with $W$ set to zero.
The limit $\lim_{k \to \infty} H_{\rm eff}(k\hat{n}) \equiv H_{\rm eff}(|\mathbf{k }| = \infty)$ is independent of the direction of the unit vector $\hat{n}$, which allows compactification of $\bk$-space to a sphere.
In practice we construct the $k=\infty$ state using independent random phases on each site.

Assuming that $H_{\rm eff}(\bk)$ is finite, gapped, and continuous for all $\bk$, this construction provides a mapping between infinite amorphous Hamiltonians and continuum Hamiltonians.
In the thermodynamic limit the effective Hamiltonian (also the effective Green's function) is invariant under continuous rotations
\begin{equation}
H_{\rm eff}(\bk) = U_{\theta} H_{\rm eff}\left(R_{\theta}^{-1} \bk \right) U_{\theta}^{-1},
\end{equation}
where $R_{\theta}$ is a real-space dimensional rotation matrix with an angle $\theta$, and $U_{\theta}$ is the on-site angular momentum representation.

We also define a related quantity, the $\bk$ and $m$-resolved spectral function
\begin{equation}
A(\bk, m, E) = \bra{\bk, m} \delta(H - E) \ket{\bk, m} = -\frac{1}{\pi} \operatorname{Im} \bra{\bk, m} G^R(E) \ket{\bk, m},
\end{equation}
where $\ket{\bk, m}$ are the states defined in \eqref{eq:plane_wave} and $G^R(E) = \lim_{\eta\to 0} \left( E + i \eta  - H\right)^{-1}$ is the retarded Green's function of the full system.
In the manuscript we show spectral functions which are summed over some $m$ values, corresponding a coarser resolution of various angular momentum characters.
For energies close to the Fermi level, the peaks of the spectral function closely follow the spectrum of the effective Hamiltonian eigenvalues, and show a band inversion across the phase transition resembling that of crystalline systems.

\end{document}